# EVIDENCES FOR SKELETAL STRUCTURES IN TORNADO AND THE PROBABLE ROLE OF NANOTUBULAR DUST IN THE ORIGIN OF TORNADO

A.B. Kukushkin, V.A. Rantsev-Kartinov
RRC "Kurchatov Institute", Moscow, 123182, Russia

**Abstract.** The results are presented of an analysis, with the help of the method of multilevel dynamical contrasting of the images, of available databases of tornado's images. This analysis extends some preliminary results on identification of skeletal structures in tornadoes (Phys. Lett. A 306 (2002) 175) and enables us to apply to the case of tornado our former hypothesis for the probable role of nanotubular dust in the origin and stability/longevity of filamentary structures of a skeletal form in electric discharges. A hypothesis for the contribution of nanotubular dust to initiation of tornadoes is suggested, in which a key role is delivered to ability of the hypothetical skeleton inside the thundercloud to provide fast long-range transport of electricity. Thus, tornado's initiation is suggested to be an electrostatic instability caused exclusively by the presence and special structuring of a nanodust.

## 1. INTRODUCTION. EXTENDING THE SKELETAL STRUCTURE PHENOMENON TO TORNADOES

Recently a phenomenon of skeletal structures (namely, those of particular topology – tubules, cartwheels, and their simplest combinations) have been found in a very wide range of length scales [1(A)], in the course of verifying the hypothesis [1(B)] which suggested the long-lived filaments, observed in a gaseous Z-pinch, to possess a skeleton which might be assembled during electric breakdown, prior to appearance of major plasma, from wildly produced carbon nanotubes (or similar nanostructures of other chemical elements). The proof-of-concept studies showed the presence of tubular and cartwheel-like structures in

(i) the high-resolution (visible-light and x-ray) images of plasma in tokamaks, Z-pinches, plasma focus, and laser-produced plasma (in the range 100 μm - 10 cm), including the images taken at electric breakdown stage of discharge in tokamak, plasma focus and vacuum spark (see the survey paper [1(C)]),

(ii) various types of dust deposits in tokamak T-10 (10 nm - 10 μm) [2(A)],

(iii) hail particles (1-10 cm), tornado ($10^3$-$10^5$ cm), and various objects in space ($10^{11}$-$10^{23}$ cm) [1(A)].

Two phenomena observed in the range $10^{-5}$-$10^{23}$ cm, namely

(i) topological identity (i.e. the similarity) of the above structures (especially, of the cartwheel as a structure of essentially non-hydrodynamic nature), and

(ii) the trend of assembling a structure from similar ones of smaller size (i.e. the trend towards self-similarity),

suggest all these skeletal structures, similarly to skeletons in the particles of dust and hail, to be basically a fractal condensed matter of particular topology of the fractal [1(A)]. Specifically, this matter may be assembled from nanotubular blocks in a way similar to that in the skeletons found in the submicron dust particles [2(B)]. The present status of the concept [1(B)] is given in the survey paper [1(D)].

Here we present the results of analyzing -- with the help of the method [1(E)] of multilevel dynamical contrasting (MDC) of the images -- the available databases of tornado's images (Sec. 2). This analysis extends some preliminary results on identification of skeletal structures in tornadoes [1(A)] and enables us to apply to the case of tornado the hypotheses [1(B)]. A hypothesis for the contribution of nanotubular dust to initiation of tornadoes is suggested, in which a key role is delivered to ability of the hypothetical skeleton inside the thundercloud to provide fast nonlocal transport of electricity (Sec. 3).

The present paper is composed of contributions by V.A. Rantsev-Kartinov (Sec. 2) and A.B. Kukushkin (Sec. 3) and has been prepared for an oral presentation [1(F)] at the 45$^{th}$ Annual meeting of American Physical Society's Division Plasma Physics (Albuquerque, NM, USA, October 2003). It has also been reported at plasma physics meetings [1(G,H)] held in the Russian Federation.

## 2. SKELETAL STRUCTURES IN TORNADOES: PHENOMENOLOGY

The major features of skeletal structures found -- with the help of the method of multilevel dynamical contrasting (MDC) of the images [1(E)] -- in the available databases of tornado's images, are as follows.

2.1 The presence of skeletal structures (tubules, cartwheels, and their simple combinations) in the main body of tornado and in its close vicinity. Similarity of these structures to those formerly found in a wide range of length scales, including the dust deposits in high-current electric discharges [2] and the hailstones [1(A)].

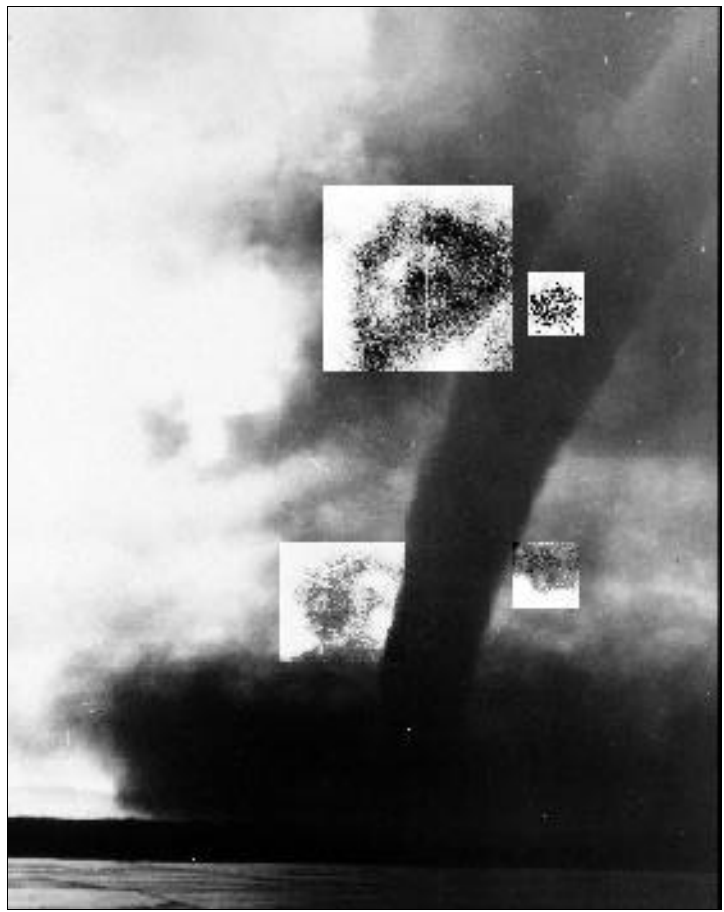

(a)

Fig. 1. (a) The image of a massive tornado (the original is taken from collection [3], http://www.photolib.noaa.gov/historic/nws/images/big/wea00216.jpg). The images in the windows around main body of tornado's column, are processed with MDC method and magnified in the following figures: (b) lower left; (c) lower right; (d), upper right; (e) upper left;.

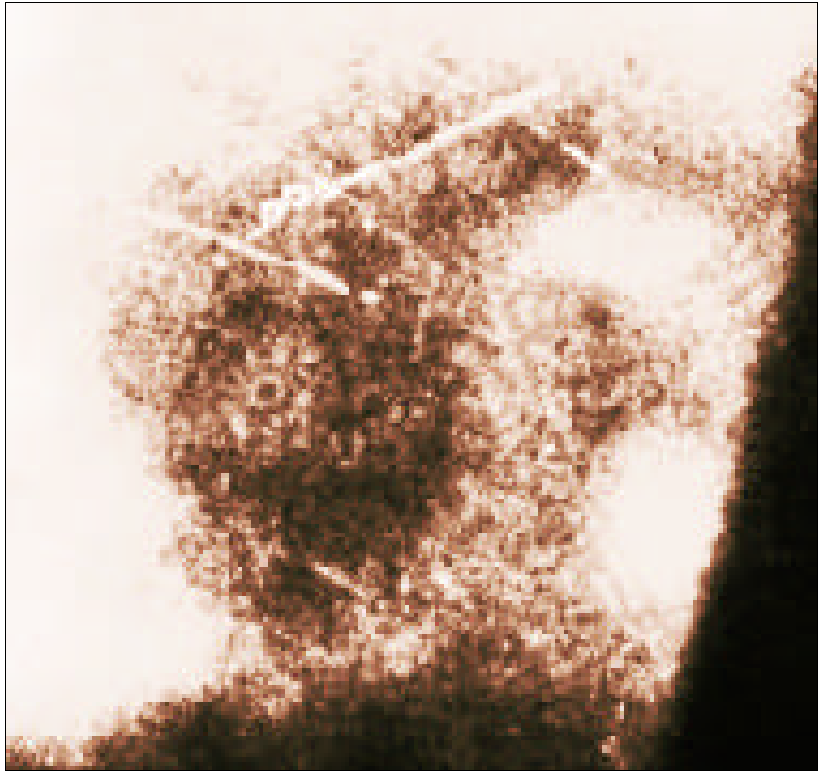

Fig. 1(b). The cartwheel-like structure is seen in the center of the image. The straight filaments may posses tubular structure as is suggested by the images of their edges.

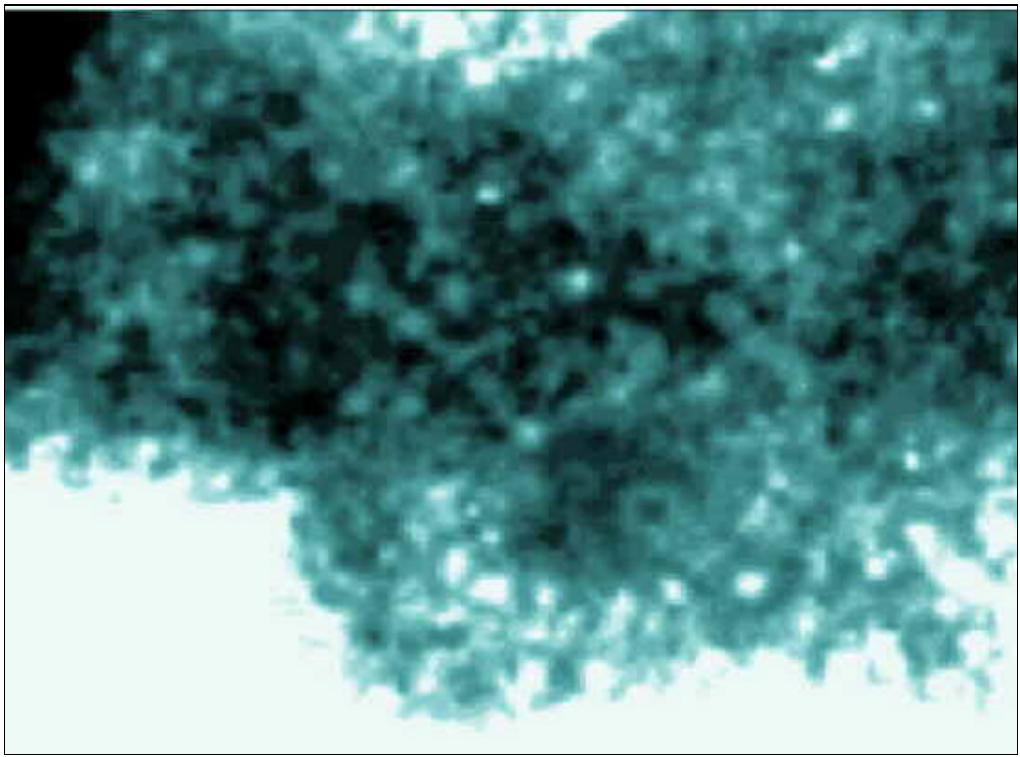

Fig. 1(c). This shows the cartwheel whose elliptic image is seen in the centre of the picture. The cartwheel is located on a long axle-tree directed downwards to the right, and is ended with a bright spot on the axle's edge.

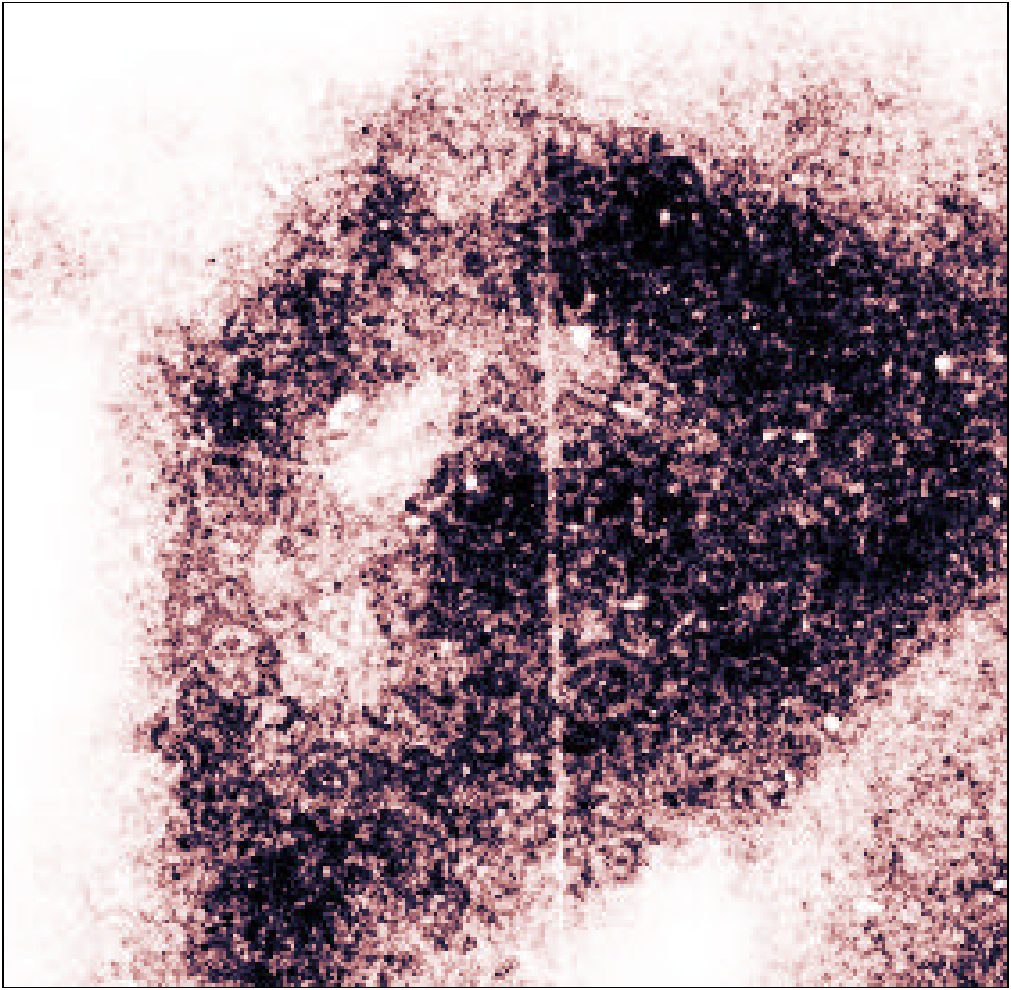

Fig. 1(d). This illustrates the presence of a complicated network of filamentary structures.

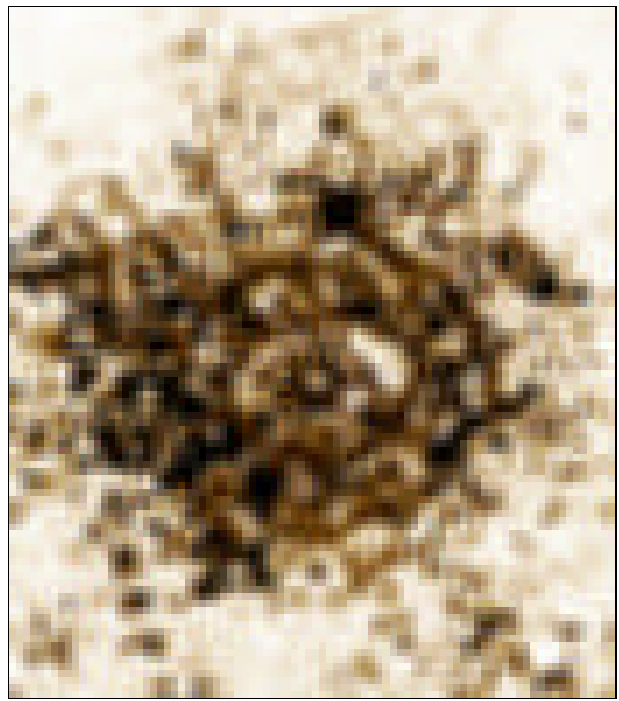

Fig. 1(e). The cartwheel-like structure with the axis directed to the observer.

2.2. General layout of skeletal structures in the main body of tornado is similar to that in a straight Z-pinch. First, the cartwheels on the axle-tree outside the main body of tornado are directed transversely to the main body – similarly to cartwheels located outside hot plasma column of the Z-pinch (cf. Fig. 1(c) and the images of a gaseous Z-pinch in [1(B,C)] ). Second, the main body of tornado possesses large tubular formations which are also directed transversely to the funnel (see Figs. 2(b), 3). Similar structuring appears in the funnel arising from the ground (see Fig. 4).

2.3. The resolvability of skeletal structures in the photos taken with large enough exposure suggests that these structures are moving slowly as compared to gas motion -- and therefore are, to a large extent, decoupled from the motion of gaseous component.

2.4. The transient region between tornadic thundercloud and the funnel appears to possess distinct skeletal structures with a complicated junction of blocks (see Figs. 2(c,d)).

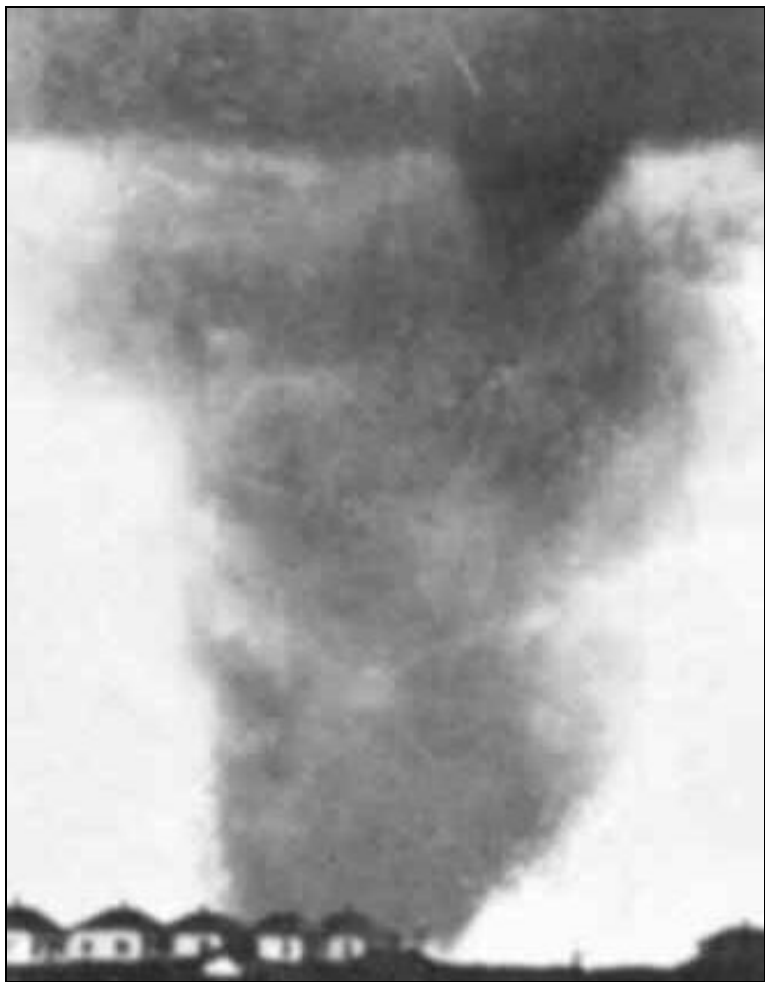

Fig. 2(a). The image of tornado at Norton, Kansas (June 1909), taken from collection [3], http://www.photolib.noaa.gov/historic/nws/images/big/wea00248.jpg.

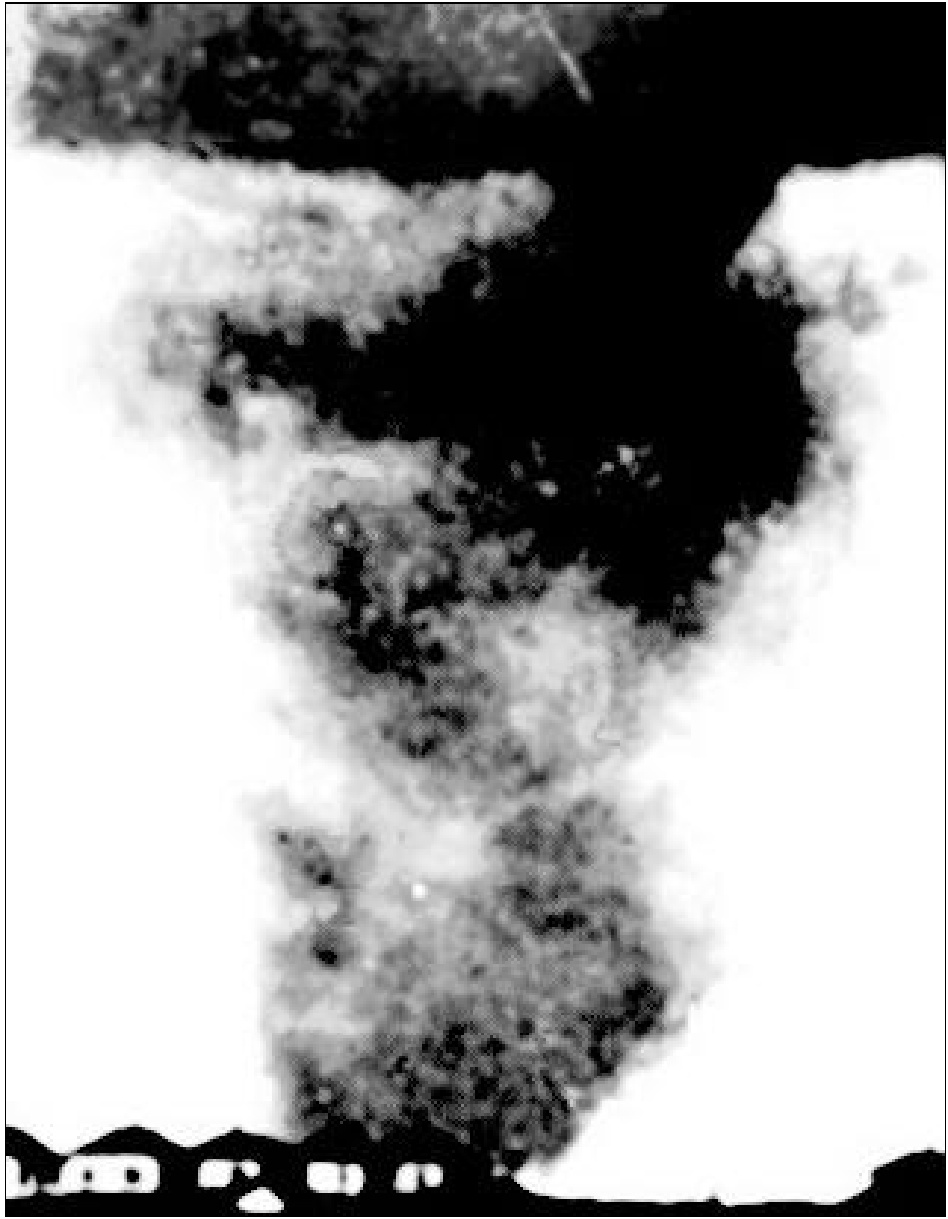

Fig. 2(b). The MDC-processed image of tornado of Fig. 2(a). The tubular formations in the main body of tornado are directed nearly horizontally and sometimes possess a resolvable coaxial structure.

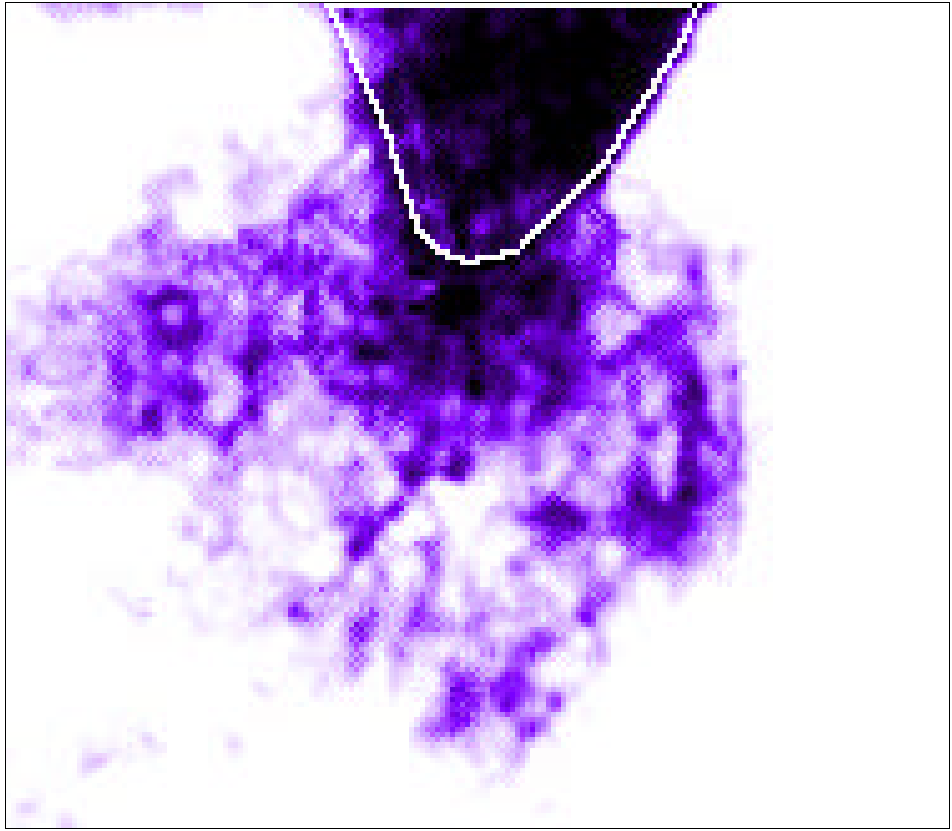

Fig. 2(c). The MDC-processed image of the fragment of the image of Fig. 2(a), namely the dark conical formation, which is seen in the right upper part of Fig. 2(a) -- the transient region between the thundercloud and the funnel. The boundary of the cone is marked with a white curve. The edge of the cone seems to be the center of tubular formations directed horizontally. One of such skeletal tubules directed to the left from the cone's edge, is seen as the parallel ellipses.

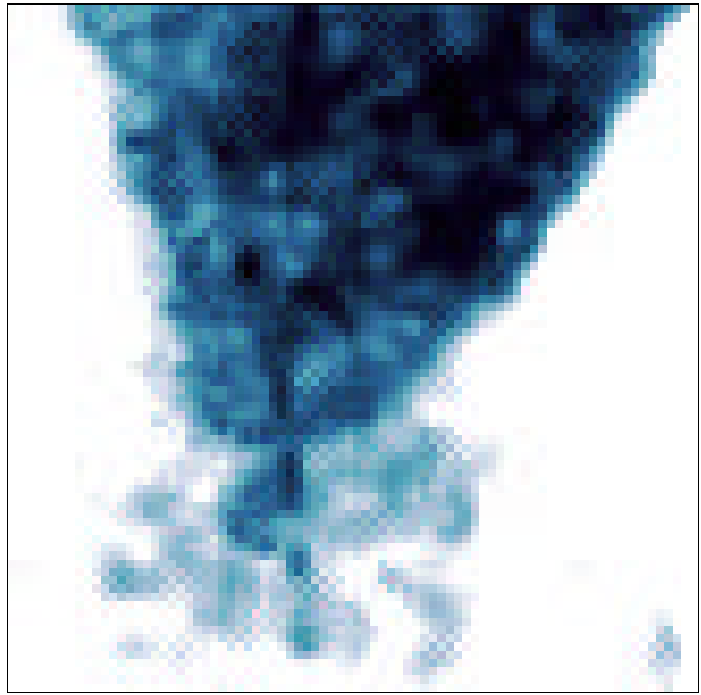

Fig. 2(d). The fragment of the image of Fig. 2(c) (namely, the edge of the cone marked with the white curve). There is a distinct vertical filament and the coaxial ring-shaped structures around this filament.

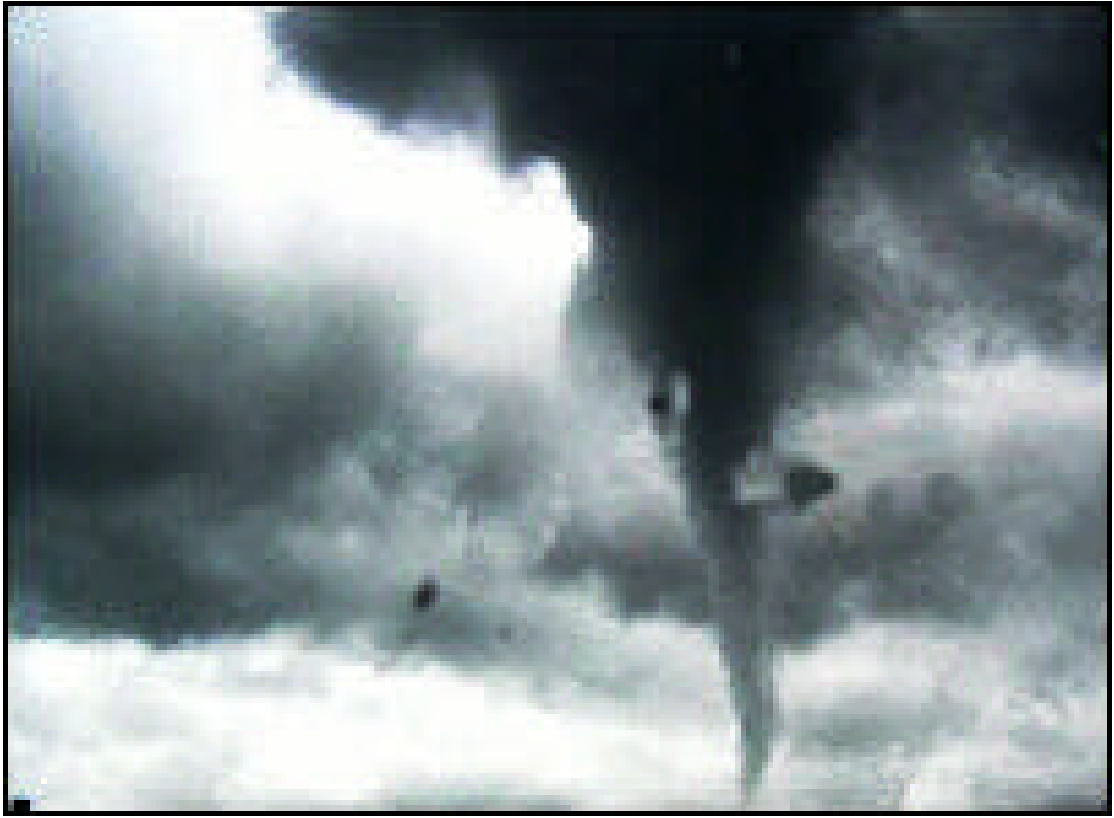

Fig. 3. The image of tornado with the distinct large horizontal branch. Interestingly, there is one more horizontal structure outside tornado' funnel (it is located to the right from the funnel), which possesses a distinct coaxial structure. The dark spots to the left from tornado may be shown to be the centers of similar tubular formations.

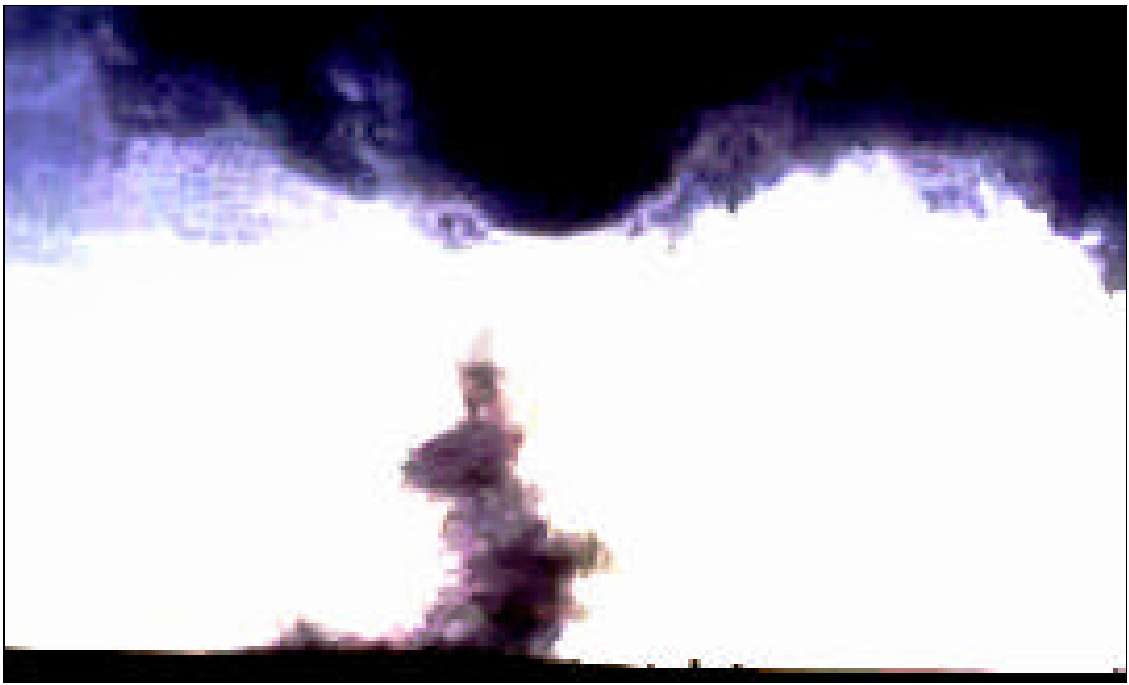

Fig. 4. This picture illustrates the presence of tubular formations directed transversely to the funnel, in the case of the funnel arising from the ground.

2.5. <u>Dendricity</u> of tornado's body [1(A)] (see Fig. 5(a)).

2.6. <u>Electric torch</u>-like structuring of tornado's funnel (see Fig. 5(a)). (Such a structuring is similar to that found in a very broad range of length scales, see [1(A)]).

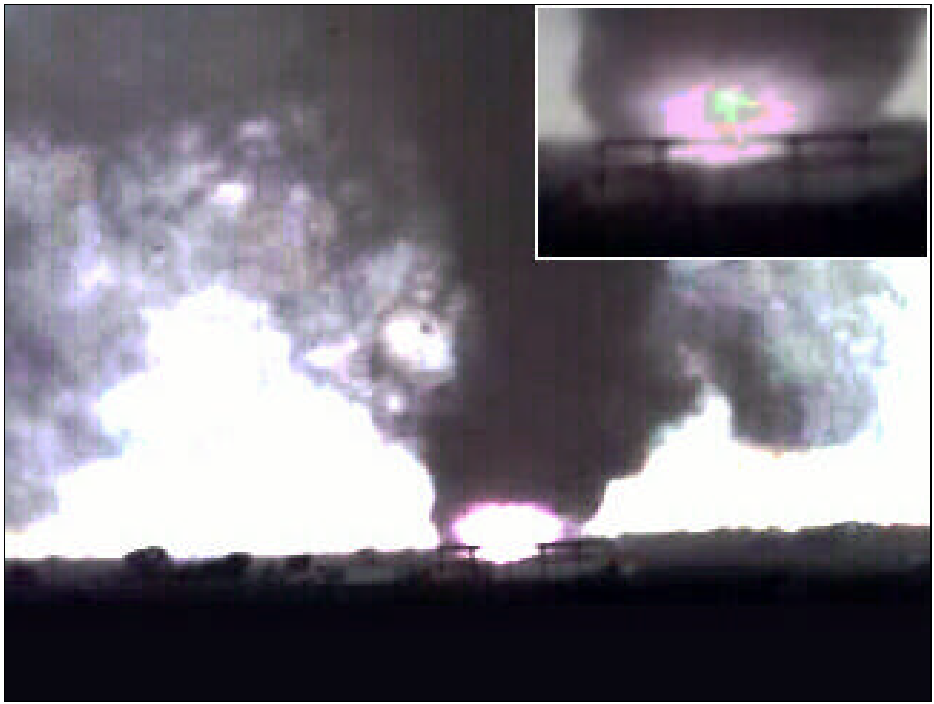

(a)

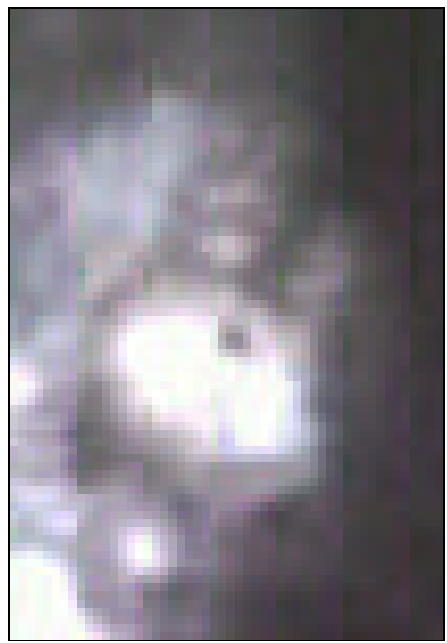

(b)

Fig. 5. (a) A fragment of the MDC-processed photographic image of a dendritic tornado with tubular funnels/branches of estimated diameter ~100 m (cf. http://www.tornadochaser.net/). It looks like there is an internal illumination inside the funnels. The edge of major funnel is shown in the insert in the right upper corner. Two other funnels do not contact the ground. The first is seen in the right-hand side of the picture as a dark branch directed opposite to the observer. The second (figure b) is located upper to the left from major funnel – it is seen like the bright edge of a light-guiding branch on the background of the opaque main body of major funnel. This branch seems to possess a tubular skeletal structure with an internal coaxial dark rod.

2.7. Low-precipitation supercell may produce an intense hailstorm with large enough hailstones.

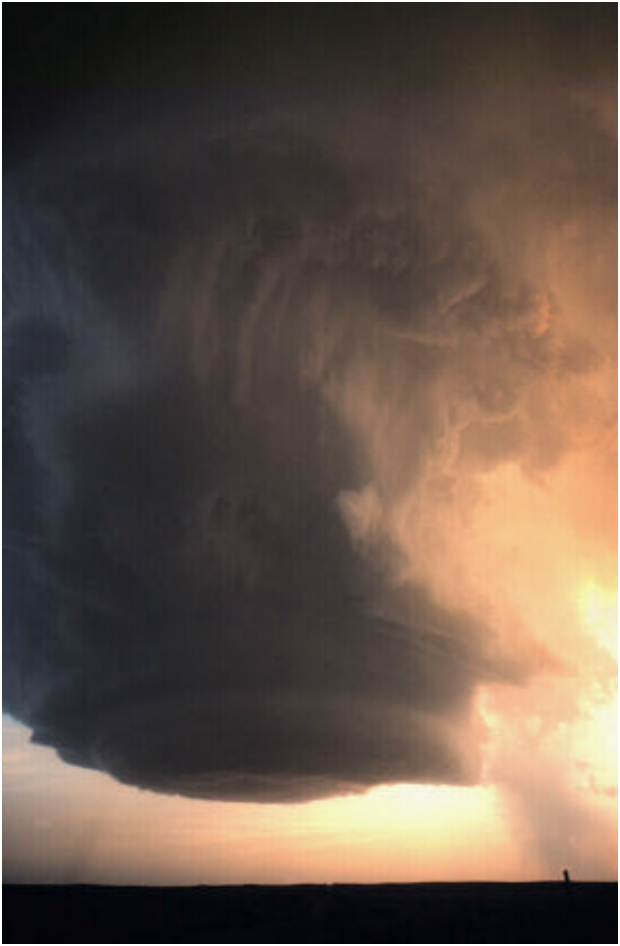

**Vertical Looking LP Supercell and Big Hail Producer**

This is an example of a LP thunderstorm in far western Oklahoma. The hail streamers were evident on this storm in advance of the main storm cloud. Much of the hail came out of the anvil, the ice crystal cloud that sweeps down stream in the jet stream. Vehicles were struck many miles in front of the storm with baseball hail while in the sunlight of the setting sun. Most did not know where the hail came from. In some cases storms like this will throw hail out the top and it may land anywhere within a few miles of the main thunderstorm cloud. This storm rotated for hours and had a tornadic signature on radar but no tornadic circulation ever made it to cloud base or the ground. Cells like this are a hazard to aviation flying in the near vicinity.

*http://www.chaseday.com/hailstorms.htm*

Fig. 6. Vertical Looking Low-Precipitation (LP) Supercell which is a Big Hail Producer (the image is taken from http://www.chaseday.com/hailstorms.htm).

2.8. The trend toward <u>self-similarity</u> in the skeletal structures. It appeared difficult to identify the trend toward self-similarity in the images of tornadoes. However, this trend is quite distinct in the hailstones (see Figs. 7(a,b,c) ).

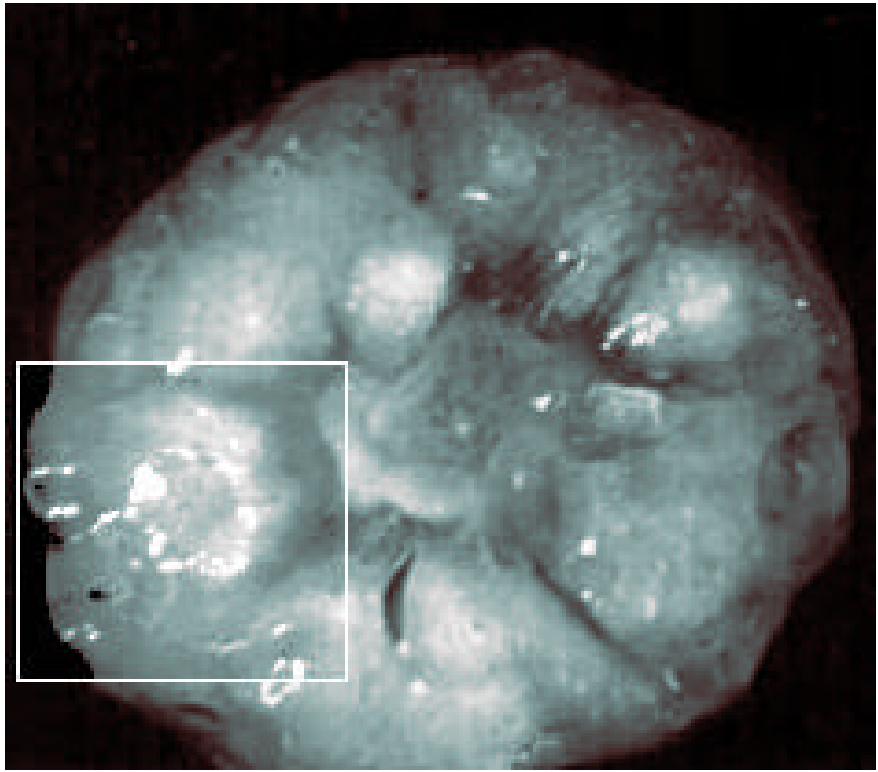

(a)

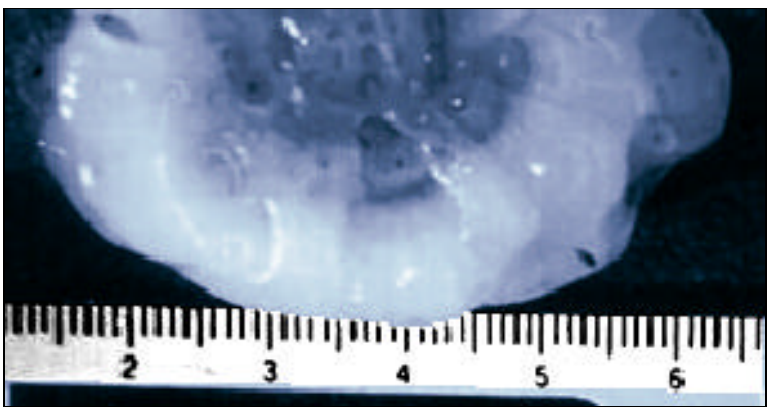

(b)

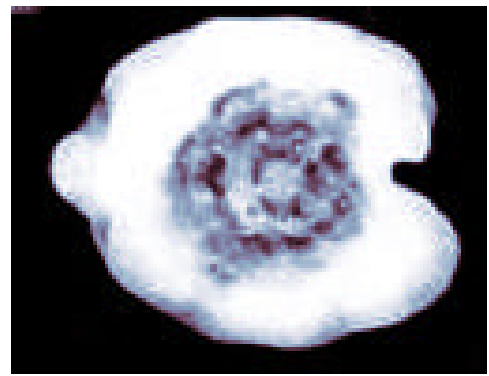

(c)

Fig. 7. The photographs [4] of hailstones of diameter 4.5 cm (a), 5 cm (b), and 5 cm (c). The frame in the left lower part of the image 'a' is MDC-processed (namely, has a higher contrast) to show an elliptic image of the edge of the radially directed tubular structure. The entire structure seems to contain a number of similar radial blocks.. Image 'b' shows strong radial bonds between the central point and the wheel. A distinct coaxial structure of the cartwheel type is seen in the central part of the image 'c'. The trend toward self-similarity is manifested itself in the presence of small cartwheels on the outer side of the entire cartwheel – namely, in the outer edges of the (slightly conical) radially-directed tubules which serve as the spokes of the entire wheel.

2.9. Sometimes hailstorm exhibits the signs of a regular structuring and presumably takes place at a low enough speed of falling dawn that allows the hailstones to survive in their collision with the ground.

2.10. The existence of areas of most frequent tornadoes (first of all, Tornado Alley in USA) may be associated with the delivery of nanodust material from volcanoes in Africa (see Fig. 8 taken from [3]).

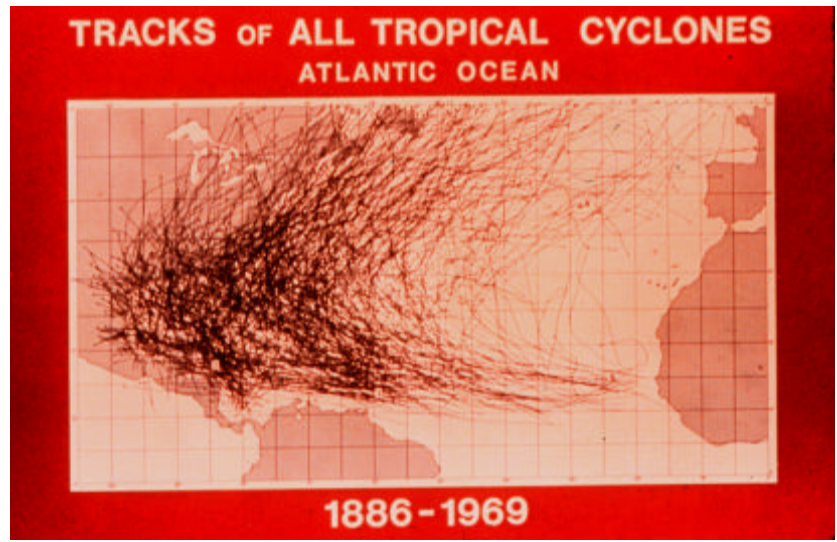

Fig. 8. (see http://www.photolib.noaa.gov/historic/nws/images/big/wea00419.jpg )

## 3. A HYPOTHESIS FOR THE ROLE OF NANODUST IN THE ORIGIN OF TORNADO

The current status of treating the tornado's mystery may be illustrated with the recent findings [5] from the project VORTEX (Verification of the Origins of Rotation in Tornadoes Experiment):
«VORTEX has produced a number of troubling new findings.

For example, it appears that perhaps *many fewer* supercells and mesocyclones produce tornadoes than scientists originally believed. At one time, researchers felt that tornadoes somehow were caused directly by mesocyclones, and that perhaps one-half of all mesocyclones produced tornadoes. We now know that this is not the case, and that tornado formation is a complicated process that depends perhaps only indirectly on the presence of a mesocyclone.

Further, we have learned that the *difference between tornadic and non-tornadic mesocyclones can be very, very subtle*. We are examining a case in which a storm shows all indications of being tornadic on WSR-88D, and in fact in mobile Doppler radar data it has a vortex with a hook and an "eye" in the hook... indicative of very strong rotation and the centrifuging of raindrops, within about 100 m of the ground. This pattern of reflectivity and velocity is in most respects the same as observed in the tornadic supercells. Yet, no tornado formed.»

The quoted valuable conclusions [5] well justify a search for the contribution of an unrecognized mechanism to initiation of tornadoes. We think it reasonable to formulate the following hypotheses for the possible contribution of nanodust (or generally speaking, of a condensed-matter component) to the origin of tornadoes.

3.1. Tornadic thundercloud may possess an internal skeleton which is hidden in the water vapor, air-based dust, etc. The skeleton may be composed of the nanodust. The nanodust particles (presumably, carbon nanotubes or similar nanostructures involving other chemical elements) in the skeleton, as it was suggested [1(B)], may be magnetically coupled by the magnetic flux trapped in the nanotubular block. Such a skeleton is flexible and restructurable, being a sort of the fractal aerogel.

The fractality of the skeleton is suggested by the total combination of the following observations (Sec. 2):

(i) similarity of skeletal structuring at different length scales (e.g., the presence of cartwheels both in tornado column and in hailstones),

(ii) self-similarity in the hailstones of cartwheel-like form (see Fig. 7),

(iii) correlation between low-precipitation supercells and intense hailstorms (see Fig. 6).

In fact, these observations show that the topology of the most rigid building blocks of the skeleton may be of a universal type.

3.2. The skeleton in tornadic thundercloud is responsible for the <u>fast long-range transport of electricity</u> (e.g., of that caused by the electric charge acquired by the skeleton during condensation of charged water drops on it) and for accumulation of large enough electric charge at certain points determined by the geometry/morphology of the skeleton (e.g., at the skeleton's edge). (Here, the transport of electricity is fast as compared to that provided by the gaseous/aerosol medium in the skeletonless thundercloud).

This implies that actually the difference between tornadic and non-tornadic thunderclouds is determined by the transport properties (in particular, by the very presence) of an internal skeleton, namely by the ability of the skeleton to collect, transport and focus the electric (and magnetic) energy. In particular, skeleton as a condensation center may substantially speed up the conversion of the latent heat into gas/plasma motion.

3.3. The localization of large enough electric charge on/around the skeleton initiates a sort of <u>electric breakdown</u> between the thundercloud and the Earth. Thus, the initial phase of tornado may be treated as an electric breakdown process which is eye-visible in the real time. Presumably, most frequently the cathode is the thundercloud, the virtual anode is on the Earth surface. Thus, tornado's initiation is suggested to be an electrostatic instability caused exclusively by the presence and special, skeletal structuring of the nanodust.

3.4. Tornado's column/funnel may be interpreted as a <u>long-lived filament</u> of (low-magnitude) electric current which is being resulted from the process of aforementioned electric breakdown. The skeleton of this filament is being build up of the restructured blocks of the skeleton in the thundercloud during a «pullout» of the parent thundercloud's skeleton downwards by the electric force which disturbs the balance of the forces in the stable, tornadoless thundercloud. (Here, longevity implies that the lifetime largely exceeds that of the lightning).

3.5. Rotation of tornado's column/funnel seems to be an implication of local restructuring of the skeletal component inside the thundercloud. That's why the rotation in the tornadic mesocyclone is actually not a source of tornado's initiation (cf. [5]). Therefore, the problem of tornado rotation may be decoupled from the problem of tornado's initiation and has to be treated in terms of the behavior of the ambient gaseous and aerosol components of the thundercloud in the presence of electric (and magnetic) field provided by the skeleton in the course of its restructuring.

3.6. The <u>early diagnostics</u> of tornado has to be aimed at identifying a skeleton inside a thundercloud and especially at those patterns of skeleton's behavior/restructuring which might be dangerous for tornado's initiation.

## 4. CONCLUSIONS

The contents of Secs. 2,3 suggest the role of electricity in tornado to be of much larger importance than it is thought by the most of severe weather researchers community (cf. wonderful comment on this issue in [6], and the references and web links therein). The universality of the phenomenon of skeletal structuring [1(A)], and the recent progress in the newest chapters of condensed-matter physics (first of all, carbon nanotubes), enable us to suggest the possibility to substantially modify the existing approaches to the probable significant role of electricity in tornadoes (cf., e.g., recent paper [7]). Tornado, because of its exceptional property of concentrating the energy density in severe weather phenomena, seems to be the best candidate for verifying both the phenomenon of skeletal structuring as itself and the hypothesis for the crucial role of a fractal condensed matter (most probably, nanotubular dust) in severe weather phenomena.